\documentstyle[12pt,aps,epsf]{revtex}

\begin{document}
\title{\vspace{2cm} {\large {\bf Two-Magnon States in {Cu(NO$_{3}$)$_{2}\cdot 2.5$D$%
_{2}$O} using Inelastic Neutron Scattering.}}}
\draft
\author{D.\ A. Tennant$^{1,2,3,4}$, C. Broholm$^{5,6}$, Daniel H. Reich$^{5}$, S.\
E. Nagler$^{4}$, G.\ E. Granroth$^{4}$, {T.~Barnes}$^{7,8}$, G. Xu$^{5},$
B.\ C. Sales$^{4}$ and Y. Chen$^{5}$.}
\address{$^1$Rutherford Appleton Laboratory, ISIS Facility, Didcot OX11 0QX, Oxon.,\\
United Kingdom\\
$^2$Oxford Physics, Clarendon Laboratory , Parks Road, Oxford OX1 3PU,\\
United Kingdom\\
$^3$Condensed Matter Physics and Chemistry Dept., Ris\o\ National Laboratory,%
\\
DK-4000, Roskilde, Denmark\\
$^4$Solid State Division, Oak Ridge National Laboratory,\\
Oak Ridge, TN 37831-6393 \\
$^5$Department of Physics and Astronomy, The Johns Hopkins University,\\
Baltimore, MD 21218 \\
$^6$National Institute of Standards and Technology, Gaithersburg, MD 20899 \\
$^7$Department of Physics and Astronomy,\\
University of Tennessee, \\
Knoxville, TN 37996-1501 \\
$^8$Physics Division, Oak Ridge National Laboratory,\\
Oak Ridge, TN 37831-6373 \\
}
\date{\today}
\maketitle

\begin{abstract}
We report measurements of the two-magnon states in a dimerized
antiferromagnetic chain material, copper nitrate ({Cu(NO$_3$)$_2\cdot2.5$D$%
_2 $O}). Using inelastic neutron scattering, we have studied the one- and
two-magnon excitation spectra in a large single crystal of this material. We
compare this new data with perturbative expansions of the alternating
Heisenberg chain and find good agreement with these calculations. The data
may also show evidence for the recently proposed $S=1$ two-magnon bound
state (Phys. Rev. B54, R9624 (1996)).
\end{abstract}


\vspace{0.8cm}

\vspace{0.8cm}

\newpage

Competition between hopping and binding effects in elementary excitations is
a general feature of low dimensional hard core systems. Such effects are
predicted to be manifest in the structure factor of multiparticle continua,
accessible using neutron and Raman scattering, as bound modes and
enhancement of continuum scattering \cite{Uhrig}\cite{Damle}. Experimental
systems of particular interest in this respect are the $S=1/2$\ alternating
Heisenberg chains (AHC) \cite{Uhrig} and the new spin ladder systems \cite
{Damle} which are predicted to have bound modes below the two-magnon
excitation continua. To investigate this phenomenon we present an
experimental study of the two-magnon states in a near ideal example of an
AHC compound, using inelastic neutron scattering.

The $S=1/2$ AHC spin Hamiltonian is 
\begin{equation}
H=\sum_{i}\ J\;{\vec{S}_{2i-1}}\cdot {\vec{S}_{2i}}+\alpha J\;{\vec{S}_{2i}}%
\cdot {\vec{S}_{2i+1}}
\end{equation}
where $J>0$ is the intradimer coupling, and $\alpha J$ ($0\leq \alpha \leq 1$%
) the interdimer one, which alternate between chain sites $i$ \cite{BRT}.
Computational methods are very effective at calculating perturbative
properties of (1), and ground and low-lying excited state wavefunctions are
given to ${\cal O}(\alpha ^{5})$ in \cite{BRT}, together with many
experimentally important quantities to ${\cal O}(\alpha ^{9})$. Higher order
expansions for selected quantities have recently been reported \cite{CCM}.
Here we compare these results to {Cu(NO$_{3}$)$_{2}\cdot 2.5$D$_{2}$O}, or
CN for short.

The magnetic properties of CN are well characterized. It is monoclinic ($%
I12/c1$ \cite{space_group}), with low temperature lattice parameters $a=16.1$%
, $b=4.9$, $c=15.8$~\AA\ and $\beta =92.9^{\circ }$\cite{IRIS}, with spin $%
S=1/2$ magnetic Cu$^{2+}$ ions. The dominant magnetic exchange integral $J$
is between pairs of spins forming dimers. Dimers separated by ${\bf u}$ are
coupled together by exchanges $J_{{\bf u}}^{\prime }$. Only the $J_{[%
{\frac12}%
,\pm 
{\frac12}%
,%
{\frac12}%
]}^{\prime }$\ exchange paths are of appreciable strength, giving two sets
of alternating Heisenberg chains (AHCs) running in the $[111]$ and $[1%
\overline{1}1]$ directions which repeat every ${\bf u}_{0}=[111]/2$ and $%
{\bf u}_{0}^{\prime }=[1\overline{1}1]/2$\ respectively.

Bulk magnetic measurements give information on the gap, exchange and ground
state energies: Applied magnetic fields induce spin flop ordering in CN
above a critical field $H_{c1}\approx 27$~kOe, with a transition to full
alignment at $H_{c2}\approx 43$~kOe. Because the orbital moment is quenched,
and demagnetization effects are negligible, $H_{c1}$ directly gives the
excitation gap, $\Delta =0.378\pm .007$~meV \cite{Diederix} and $H_{c2}$
gives the sum of exchange couplings $J+\sum_{{\bf u}}J_{{\bf u}}^{\prime
}=0.580\pm .007$~meV. Normally it is not possible to measure the ground
state energy of quantum antiferromagnets but high field magnetization
techniques in our case make this achievable.

The ground state energy-per-spin $e_{0}$ can be found from the low
temperature isothermal magnetization $M(H)$ using $e_{0}\approx
e_{f}-SH_{c2}+\int_{0}^{H_{c2}}M(H)dH$ where the fully aligned
energy-per-spin is $e_{f}=S^{2}/2\cdot (J+\sum_{{\bf u}}J_{{\bf u}}^{\prime
})=H_{c2}/8=0.0725\pm .001$~meV. Using the 270 mK data of Diederix {\it et
al.} in Figure (3) of \cite{Diederix} (measured using proton resonance) to
determine the integral over magnetization gives an experimental ground state
energy-per-spin $e_{0}=-0.174\pm .004$~meV. This is essentially the $T=0$
result, as the gap activation energy corresponds to 4.4 K. To estimate
thermodynamic properties we approximate the sum of the exchanges by the
single coupling $\alpha J=\sum_{{\bf u}}J_{{\bf u}}^{\prime }$ of equation
(1). 
Using the ${\cal O}(\alpha ^{9})$ expansions \cite{BRT} for $\Delta (\alpha
) $ and $e_{0}$ gives $J=0.455\pm .002$~meV and $\alpha =0.277\pm .006$; in
agreement with the results of \cite{Diederix} and \cite{Bonner}, $J=0.45$
meV and $\alpha =0.27$. Our calculated values of the thermodynamic
parameters $J+\sum_{{\bf u}}J_{{\bf u}}^{\prime }=0.581$ meV, $\Delta =0.379$
meV and $e_{0}=-0.172$~meV agree within error with the experimental values.

The neutron scattering structure factor ${\cal S}({\bf Q},\omega )$ probes
the ground $|G\rangle $ and excited states $|E\rangle $ of a magnetic system
through the matrix element $|\langle E|S^{\alpha }({\bf Q})|G\rangle |^{2}$
of the Fourier transformed spin operator $S^{\alpha }({\bf Q})$ $(\alpha
=x,y,z)$. We measured ${\cal S}({\bf Q},\omega )$ using inelastic neutron
scattering from two deuterated single crystals of CN with a total mass of
14.1 g, using the SPINS cold neutron triple-axis spectrometer at the NIST
Center for Neutron Research. The substitution of D for H reduces incoherent
scattering of neutrons and does not significantly change the magnetic
properties of the material. The sample was mounted with $(h,0,l)$ as the
scattering plane in a pumped $^{3}$He cryostat at a base temperature of
300~mK.

The spectrometer was set up with $80^{\prime }$ before the sample as the
only collimation. A vertically focused pyrolytic graphite PG(002)
monochromator and a horizontally focused analyzer array composed of eleven
independently rotatable PG(002) blades were employed. A cooled Be filter
before the sample removed higher-order contamination from the beam.
Measurements were made with fixed final energy $E_{f}=2.5$~meV by scanning
incident energy at various reduced wavevector transfers along the chain, $%
\widetilde{q}={\bf Q}\cdot {\bf u}_{0}$. The wide angular acceptance ($%
14^{\circ }$) of the analyzer dominated the instrumental resolution making
it highly elongated perpendicular to the scattered wavevector in the
scattering plane. Scan trajectories were chosen to maintain the final
wavevector ${\bf k}_{f}$ along the $(101)$ direction so as to integrate over
nondispersive directions while maintaining good resolution in $\widetilde{q}$%
.

Because the ground state is a singlet it cannot be probed directly by
neutrons, but its composition is reflected through spin matrix elements to
triplet excited states. Neutron scattering matrix elements to the $S=1$ one
magnon states have been calculated to ${\cal O}(\alpha ^{5})$ \cite{BRT}:
the leading order scattering process is from the bare dimer component of the
ground state, and an $\alpha /2\cdot \cos (\widetilde{q})$ component in the
one magnon structure factor arises from an $O(\alpha )$ two-dimer excitation
in the ground state. Similarly, transitions between various components of
the full ground and excited states can be identified through distinctive
modulations of the ${\bf Q}$ dependence of ${\cal S}({\bf Q},\omega )$.

Figure (1) shows scans in energy in CN: Panel (a) shows a scan at the
antiferromagnetic zone-center, $\widetilde{q}=2\pi $, taken at $T=300$~mK.
Strong elastic scattering from incoherent nuclear processes is clearly seen
as well as a one magnon \cite{IRIS} peak at 0.4~meV, close to the dimer
energy $J=0.45$~meV. The non-magnetic background (dashed line) was modelled
by a Gaussian (incoherent) component and a power-times-Lorentzian (broad,
quasielastic) component. A second magnetic peak appears at about 0.9~meV,
double the dimer energy. Panels (b) and (c) show this peak with the
nonmagnetic background subtracted at $\widetilde{q}=2\pi $ and $\widetilde{q}%
=3\pi $, respectively. This feature is considerably weaker than the
one-magnon scattering and narrows at the zone-boundary, panel (c); this
behavior is consistent with two-magnon scattering.

Figure (2) shows excitations calculated to ${\cal O}(\alpha )$ using
degenerate perturbation theory, including states up to two-magnon, for $%
J=0.45$ meV and $\alpha =0.27$. The calculation is similar to that for the
two-soliton continuum in the XXZ Ising chain in \cite{Ishimura}, except that
a dimer excited state basis is used and the one magnon band is included \cite
{Theor1}. An interesting feature of the spectrum is the existence of an $S=1$
two-magnon bound state for a range of $\widetilde{q}$ around $\widetilde{q}%
\approx (2n+1)\pi $ where $n$ is an integer. This bound state is due to the
attraction between adjacent excited dimers with total spin 1 (and 0). It
exists only over a limited range of $\widetilde{q}$ around the bandwidth
minimum in the two-magnon continuum because of the nature of the hopping
matrix elements. Near the two-magnon bandwidth minimum the hopping element
is small, and binding occurs. Far from this minimum the gain in hopping
energy of excited dimers is dominant and no $S=1$ bound mode exists. An
interesting consequence of the competition is that an $S=0$ bound mode (not
visible to neutron scattering, but can be observed using Raman scattering)
should exist below the continuum for all $\widetilde{q}$, because it has a
larger attractive interaction between adjacent dimers.

The strength of production of the $S=1$ bound mode in neutron scattering
depends on the spin operator matrix element to this state. Perturbation
analysis in $\alpha $ \cite{BRT} shows that there is no zeroth order
coupling of the bare dimer ground state to the two-magnon bound state and
continuum through the neutron scattering matrix element. The leading
perturbative contribution to ${\cal S}({\bf Q},\omega )$ appears at ${\cal O}%
(\alpha ^{2})$, and is due to a transition from the ${\cal O}(\alpha )$
two-excited-dimer component of the ground state and to the ${\cal O}(\alpha
) $ one-excited-dimer component of the bound state. The matrix element of
these two basis components has a complicated and characteristic wavevector
dependence.

Figure (2) shows the energy corresponding to a weighted average of
scattering as a thick grey line. It is notable that at $\widetilde{q}=3\pi $
to a good approximation the neutrons couple only to the bound mode, so that
nearly all the scattering weight is in it, not the continuum. The calculated
neutron scattering intensity from the bound state is 2\% of the one-magnon
intensity which agrees with the data in Figure (1).

The one- and two-magnon scattering at 300~mK was scanned from $\widetilde{q}%
=\pi $ to $5\pi $ in steps of $\pi /4$. The background subtracted data are
plotted in the upper panel of Figure (3). A calculation of the magnetic
scattering based on the perturbation calculation described above, including
the dimer envelope function \cite{BRT}, is shown in the lower panel of
Figure (3). The calculation is directly comparable with the data in the
upper panel of Figure (3), and at a qualitative level there is good
agreement with experiment.

A quantitative comparison between theory and data is shown in Figure (4).
The measured positions of one- and two-magnon peaks are plotted in the left
panel. Energies, widths and intensities for each peak were extracted by
least-squares fitting of Gaussians. Considerable dispersion of the
one-magnon modes is evident. Measurements of the one magnon dispersion were
previously fitted using an ${\cal O}(\alpha )$ model of chains with an
additional weak interdimer coupling $J_{[1/200]}^{\prime }$ and $%
J_{[001/2]}^{\prime }$ \cite{IRIS} and \cite{stone}.

The dispersion relation predicted by this model gives a good account of our
data, Figure (4). The two-magnon peak is broader than experimental
resolution, and the extracted positions (grey filled circles) are nearly
dispersionless and are located at the calculated weighted average energies
(grey band) replotted from Figure (2). Interchain effects are effectively
integrated over in the two-magnon scattering, and this results in line
broadening rather than shifts in energy; we thus have implicitly included
interchain coupling effects in our definition of $\alpha =0.27$ for (1).

The right-hand panels of Figure (4) show one and two magnon integrated
intensities extracted from the fit. The one-magnon intensity (lower panel)
is well described by the theory (solid line), and shows a simple
oscillation. Because the one-magnon intensity is modulated by the dimer
envelope function in ${\cal S}({\bf Q},\omega )$ \cite{BRT} the intensity
should go to zero near $\widetilde{q}=4\pi $. The residual intensity comes
from secondary elastic scattering from incoherent processes, and the theory
(solid line) is corrected for this. The two-magnon intensity (upper panel)
shows a more complicated $\widetilde{q}$ dependence, which is due to the
various basis transitions that contribute to the coupling of the ground to
excited states. Although the ${\cal O}(\alpha )$ result looks qualitatively
similar to the data, it underestimates the scattering at $\widetilde{q}%
\approx \frac{9}{2}\pi $ and overestimates it at $2\pi $, which may indicate
that higher order terms in the scattering amplitude are important. It is
notable that the two-magnon intensity is very strongly dependent on the
spatial arrangement of magnetic ions.

The binding energy of the $S=1$ state is predicted to be \cite{BRT} $%
E_{B}=J\left( \frac{1}{4}\alpha -\frac{13}{32}\alpha ^{2}\right) =0.017$ meV
for CN. Scattering around $\widetilde{q}=3\pi $ is centered\ at $0.852\pm
.007$ meV, which gives $E_{B}=0.03\pm .02$ meV. Although the energy and
intensity around $\widetilde{q}=3\pi $ lend support to binding around this
bandwidth minimum, the experimental error means this does not constitute
definitive proof of the effect in CN. The much better energy discrimination
of time-of-flight (TOF) neutron spectrometers could provide this by
resolving the bound mode from the continuum. Previously, TOF techniques have
proven successful in the study of similar binding effects at the bandwidth
minimum of the two-soliton continuum scattering of the $S=1/2$\ XXZ Ising
chain material CsCoCl$_{3}$ \cite{Goff}. Unlike the AHC, binding only occurs
when extra terms in the Hamiltonian, such as exchange mixing \cite{Goff} or
next-nearest neighbor coupling \cite{Matsubara}, are present. However
limited neutron fluxes may make such measurements difficult for CN \cite
{IRIS}.

In conclusion, we have used inelastic neutron scattering to investigate the
ground and excited states of the near-ideal alternating Heisenberg chain
material Cu(NO$_{3}$)$_{2}\cdot 2.5$D$_{2}$O. Our measurements are
consistent with predictions of this model for several magnetic properties of
this system, including the ground state energy, one- and two-magnon
excitation spectra and intensities, and possibly the existence of a
two-magnon bound state. Much experimental work remains to be done to
establish the phenomenology of binding in isotropic 1D systems.

DAT wishes to thank Drs B. Lebech, R. Hazell, and D. McMorrow for their help
and Ris\o\ for generous support. This work was partly supported by Oak Ridge
National Laboratory, managed by UT-Battelle, LLC, for the US Dept. of Energy
under contract DE-AC05-00OR22725. The NSF supported work at SPINS through
DMR-9423101 and work at JHU through DMR-9453362 and DMR-9801742. DHR
acknowledges the generous support of the David and Lucile Packard Foundation.

\begin{figure}[tbp]
\caption{ (a) Low temperature scattering at $\widetilde{q}=2\protect\pi $.
The dashed line is a fitted background and the solid line is a fit to the
scattering described in the text. (b) Two magnon scattering with background
subtracted off. The solid line is a fit (see text). The solid bar indicates
the instrumental resolution. (c) Two-magnon scattering for $\widetilde{q}=3%
\protect\pi $ with nonmagnetic background subtracted off. }
\end{figure}

\begin{figure}[tbp]
\caption{Schematic showing predictions of perturbation theory specialized to 
$J=0.45$ meV and $\protect\alpha=0.27$. }
\end{figure}

\begin{figure}[tbp]
\caption{ (Color) Upper panel shows a color filled contour plot of the
measured data with nonmagnetic background subtracted. Intensity is on a
linear scale indicated by color, going from dark red (minimum) to light
yellow (maximum). The two magnon scattering has been enhanced by a factor of 
$10^{2}$ to make it visible on the same scale. Lower panel shows the
calculated scattering using perturbation theory with corrections for
instrumental resolution, multiple scattering and magnetic form factor.}
\end{figure}

\begin{figure}[tbp]
\caption{ Comparison of theory and data. Left panel shows fitted positions
of observed scattering with predictions using perturbation theory (see
text). Right lower panel shows fitted one magnon intensity compared with
perturbation theory (see text). Right upper panel shows a comparison of
two-magnon intensity with perturbation theory (see text). }
\end{figure}


\begin{references}
\bibitem{space_group}  Note that the space group for Cu(NO$_{3}$)$_{2}\cdot
2.5$H$_{2}$O has been incorrectly given in previous work as I 1 2/a 1.
\end{references}
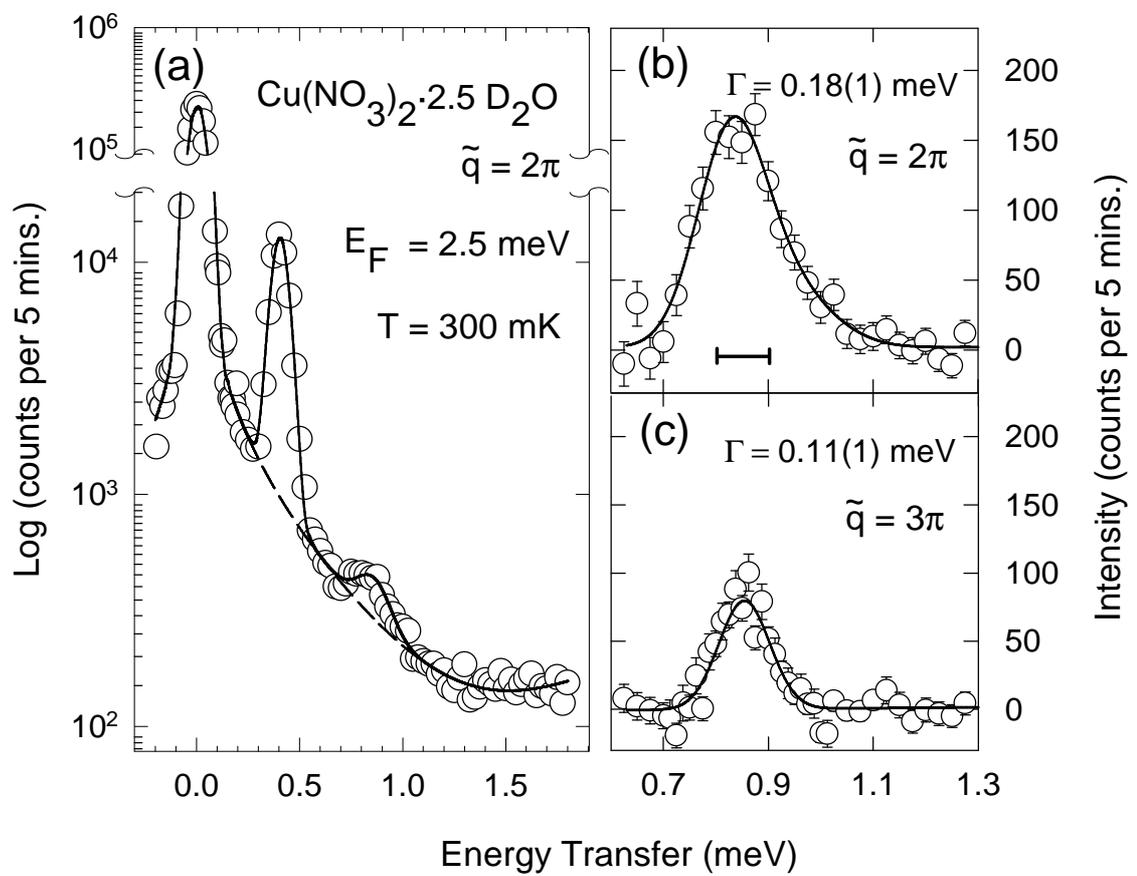
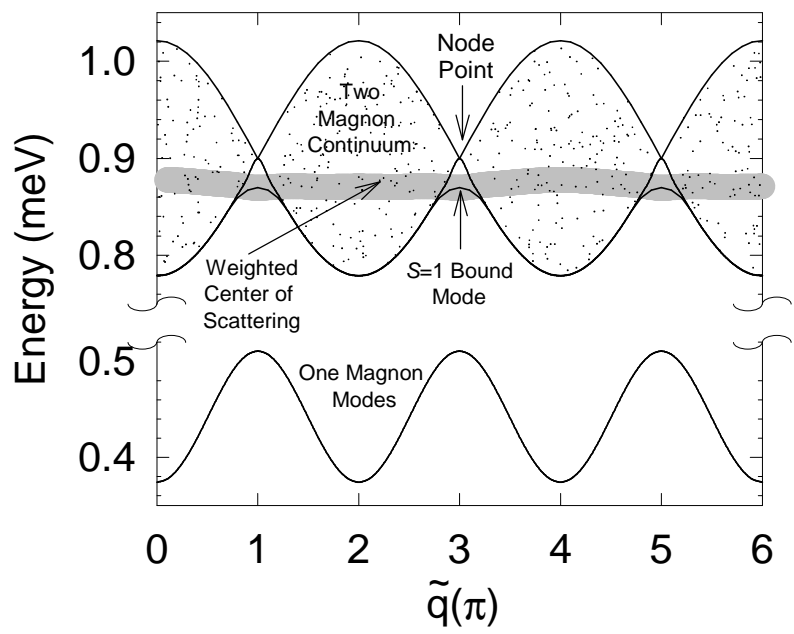
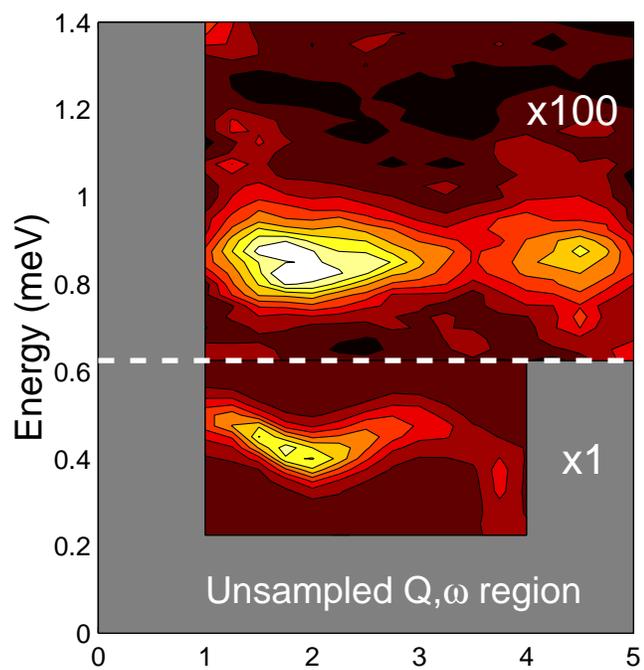
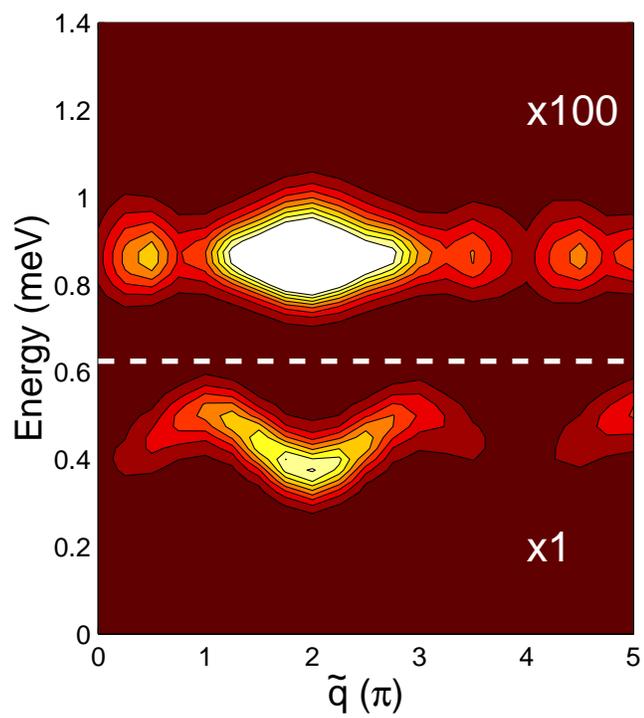
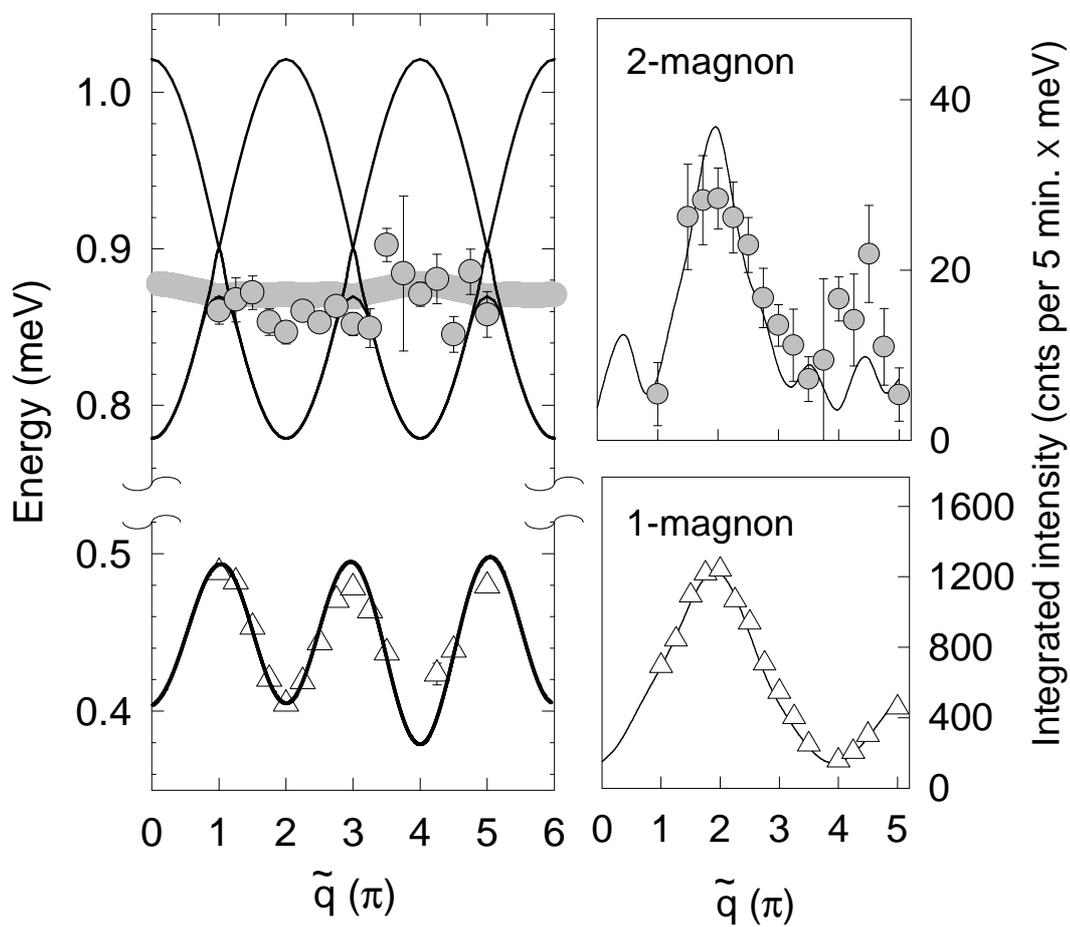
\end{document}